\documentclass[twocolumn,showpacs,prb,amsmath,amssymb,floatfix]{revtex4}
\usepackage{amsmath,amssymb}
\usepackage[latin1]{inputenc}
\usepackage{bm}
\usepackage{epsfig}
\usepackage{color, graphicx}
\usepackage{psfrag}

\begin{document}

\title{Spectral function of the
Anderson impurity model at finite temperatures}

\author{Aldo Isidori}  
%\affiliation{Institut f\"{u}r Theoretische Physik, Universit\"{a}t
%  Frankfurt,  Max-von-Laue Stra{\ss}e 1, 60438 Frankfurt, Germany}

\author{David Roosen}  
%\affiliation{Institut f\"{u}r Theoretische Physik, Universit\"{a}t
%  Frankfurt,  Max-von-Laue Stra{\ss}e 1, 60438 Frankfurt, Germany}

%\author{Hermann Freire}
%\affiliation{Instituto de Física, Universidade Federal de Goiás, 74.001-970, Goiânia-GO, Brasil}

\author{Lorenz Bartosch}  
%\affiliation{Institut f\"{u}r Theoretische Physik, Universit\"{a}t
%  Frankfurt,  Max-von-Laue Stra{\ss}e 1, 60438 Frankfurt, Germany}

\author{Walter Hofstetter}  
%\affiliation{Institut f\"{u}r Theoretische Physik, Universit\"{a}t
%  Frankfurt,  Max-von-Laue Stra{\ss}e 1, 60438 Frankfurt, Germany}

%\author{Hofstetter group} 
%\affiliation{Institut f\"{u}r Theoretische Physik, Universit\"{a}t
%  Frankfurt,  Max-von-Laue Stra{\ss}e 1, 60438 Frankfurt, Germany}

\author{Peter Kopietz} 
\affiliation{Institut f\"{u}r Theoretische Physik, Universit\"{a}t
  Frankfurt,  Max-von-Laue Stra{\ss}e 1, 60438 Frankfurt, Germany}

% \date{\today}
\date{March 19, 2010}

\begin{abstract}

Using the functional renormalization group (FRG) and the numerical renormalization group (NRG), 
we calculate the spectral function of the Anderson impurity model at zero and finite temperatures. 
In our FRG scheme spin fluctuations are treated non-perturbatively via a suitable Hubbard-Stratonovich field, 
but vertex corrections are neglected. 
A comparison with our highly accurate NRG results shows that this FRG scheme gives a  
quantitatively good description of the spectral line-shape 
at zero and finite temperatures both in the weak and strong coupling regimes, 
although at zero temperature the FRG is not able to reproduce 
the known exponential narrowing of the Kondo resonance at strong coupling.

\end{abstract}

\pacs{72.15.Qm, 71.27.+a, 71.10.Pm}

\maketitle

\section{Introduction}
In the past decade, the experimental realization of
man-made nanostructures such as quantum dots coupled to 
a metallic environment\cite{Goldhaber98, Cronenwett98} 
has brought renewed attention to the theoretical handling 
of the Anderson impurity model (AIM).
Moreover, the solution of the AIM represents one of the fundamental steps in
the so-called dynamical mean-field theory, \cite{Georges96} describing the
quantum dynamics of the Hubbard model in the limit of
infinite spatial dimensions, \cite{Metzner89} where the 
effects of the dynamical Weiss field can be described in terms of 
an effective local impurity model subject to a self-consistency condition. 
Because the thermodynamics of the AIM can be obtained exactly
by means of the Bethe Ansatz \cite{Tsvelick83,Hewson93} and, on the other hand,
Wilson's numerical renormalization
group \cite{Wilson75} (NRG) gives a numerically controlled method of calculating
spectral properties, \cite{Costi94,Hofstetter00,Peters06,Weichselbaum07,Bulla08} the AIM can also be used
as a benchmark model for testing various non-perturbative many-body methods.
It is therefore important to develop reliable methods for solving
the AIM especially at intermediate and strong coupling 
which require low computational effort, in the perspective
of tackling the more complex impurity problems arising from both
dynamical mean-field theory applications and experimental 
realizations of nanostructure devices.

Motivated by the above considerations, several authors 
have tried to reproduce at least some aspects of the
known properties of the AIM using functional renormalization group (FRG)
methods. \cite{Hedden04,Karrasch08,Bartosch09,Jakobs09}
Although the FRG gives a formally exact
hierarchy of integro-differential equations for all irreducible
vertices, \cite{Berges02,Pawlowski07,Kopietz10,Rosten10} in practice this hierarchy has to be truncated
in order to extract physical information from it.
Unfortunately, at this point no truncation of this hierarchy has been found
which correctly reproduces all the known strong-coupling properties of the AIM, such 
as the correct interaction dependence of the  Kondo scale, which is known to be
exponentially small in the interaction strength.
In fact, the proper construction of unbiased truncations of the formally exact  
FRG flow equations which do not break down 
in the strong coupling limit is a largely unsolved problem
in the field. 

Quite generally, some progress in solving the FRG flow equations 
can be made if the dominant fluctuation channel is 
known a priori. For example, for the description
of superfluidity in the attractive Fermi gas
the particle-particle channel is known to play a special role. In such a situation it is natural
to decouple the interaction in the particle-particle channel 
using a Hubbard-Stratonovich transformation and consider the FRG flow
of the coupled mixed fermion-boson model. \cite{Kopietz10,Bartosch09b,Schuetz05, Floerchinger08b,Strack08,Diehl09}

In Ref.~[\onlinecite{Bartosch09}] a similar strategy has been applied to 
the AIM, although in this case the proper choice of the Hubbard-Stratonovich decoupling
is not so obvious. For simplicity, we shall focus in this work 
on the local moment regime of the symmetric AIM 
at intermediate to strong coupling; in this case the 
physics is dominated by spin fluctuations, 
so that one should decouple the on-site interaction in the
spin-fluctuation channels.
Still, the decoupling is ambiguous because of the freedom of distributing
the interaction between transverse and longitudinal spin channels.
In the simplest case, the decoupling is then performed such that only
transverse spin-fluctuations are explicitly introduced. 
The resulting FRG flow equations have already been
derived in Ref.~[\onlinecite{Bartosch09}], but further approximations have been made
to reduce the resulting integro-differential equation to an ordinary differential equation
for the wave-function renormalization factor $Z$. 
Unfortunately, in this approximation only the low-energy
features of the spectral function can be described.
In this work we shall not rely on such a low-energy approximation,
but present a fully self-consistent solution of the
integro-differential equation for the self-energy of the AIM.
This enables us to calculate the full spectral line-shape of the
AIM at all energy scales. 
In order to test the accuracy of our FRG approach, we shall also calculate 
the spectral function at zero and finite temperatures using the
highly reliable NRG approach. By comparing the results obtained with the two 
methods, we then show that our rather simple truncation of the
FRG flow equations gives a quantitatively accurate description of the 
spectral function of the AIM at all energies.

%Moreover, by direct comparison with  the exact line-shape obtained within the
%numerical renormalization group at finite temperature, we show that
%for temperatures larger than the Kondo temperature
%our FRG approach gives  quantitatively accurate 
%results for the spectral line-shape at all energies ({\bf{hopefully, to be checked!}}).

\section{FRG with partial bosonization
in the spin singlet channel}

The AIM describes a single localized  electron level which is coupled to a band of non-interacting
conduction electrons. The latter degrees of freedom can be integrated out so that the
partition function can be written as a fermionic functional
integral involving only Grassmann fields $d_{\sigma} ( \tau )$ and
$\bar{d}_{\sigma} ( \tau )$ associated with the localized electrons.
Here $\sigma = \uparrow, \downarrow$ labels the spin projection,
and $\tau$ is the imaginary time.
Following Ref.~[\onlinecite{Bartosch09}]
we decouple the local on-site interaction
by means of a complex bosonic  Hubbard-Stratonovich field
$\chi$ in the spin-singlet particle-hole channel. The partition function
$\cal{Z}$ divided by the corresponding partition function $\cal{Z}_{\rm HF}$
in Hartree-Fock approximation can then be written as
\begin{equation}
 \frac{ \cal{Z} }{{\cal{Z}}_{\rm HF}} = \frac{ \int {\cal{D}} [ \bar{d}, d , \bar{\chi} , \chi  ] 
e^{ - S_0 [ \bar{d}, d , \bar{\chi}, \chi  ]  - 
 S_1 [  \bar{d}, d , \bar{\chi}, \chi   ] }}{  \int {\cal{D}} [\bar{d}, d , \bar{\chi} , \chi   ] 
e^{ - S_0 [\bar{d}, d , \bar{\chi}, \chi   ]  } }.
 \label{eq:Zratio2}
 \end{equation}
The Gaussian part of the bare action is
 \begin{eqnarray}
 S_0 [    \bar{d}, d , \bar{\chi} , \chi    ] & = &
- \int_{ \omega} \sum_{  \sigma  }
  {G}_0^{-1} ( i \omega ) 
  \bar{d}_{ \omega \sigma}  {d}_{ \omega \sigma}  
 \nonumber
 \\
 &  & + 
 \int_{ \bar{\omega}} 
U^{-1} \bar{\chi}_{  \bar{\omega}} \chi_{  \bar{\omega}} ,
 \label{eq:S0Phi}
\end{eqnarray}
while the interaction part can be written as
 \begin{equation}
 S_1 [   \bar{d}, d , \bar{\chi} , \chi      ]  =  \int_{\bar{\omega}} \int_{\omega}
\left[
 \bar{d}_{ \omega + \bar{\omega} \uparrow} d_{ \omega  \downarrow}
{\chi}_{ \bar{\omega}} + 
    \bar{d}_{ \omega \downarrow} d_{ \omega + 
\bar{\omega} \uparrow}
 \bar{\chi}_{ \bar{\omega}}  \right],
\label{eq:S1Phi}
 \end{equation}
where $\int_{\omega} = \beta^{-1} \sum_{\omega}$ denotes summation over fermionic
Matsubara frequencies $i \omega$, with $\beta = 1/T$ being the inverse temperature, whereas
$\int_{ \bar{\omega} } = \beta^{-1} \sum_{ \bar{\omega} }$
denotes summation over bosonic Matsubara frequencies $i \bar{\omega}$ (in the zero temperature limit 
the sums are replaced by integrals). 
For simplicity we consider only the symmetric AIM in the wide band limit, 
where the Hartree-Fock propagator is given by
 \begin{equation}
 G_0 ( i \omega ) = \frac{1}{ i \omega + i \Delta {\, \rm sign \,} \omega }.
 \end{equation}
In the above expression the energy scale $\Delta$ arises from the hybridization 
between the localized $d$-electrons, represented by the Grassmann variables
$d_{\sigma}$ and $\bar{d}_{\sigma}$, and the conduction electrons.  
The exact $d$-electron propagator, on the other hand, 
is of the form
 \begin{equation}
 G ( i \omega ) = \frac{1}{ i \omega + i \Delta {\, \rm sign \,} \omega - \Sigma ( i \omega )}.
 \end{equation}
Our aim is to calculate the irreducible self-energy 
$\Sigma ( i \omega )$ using the FRG and then analytically continue it to the real frequency axis.

We use here the frequency transfer cutoff scheme 
proposed in Ref.~[\onlinecite{Bartosch09}], where an infrared cutoff is introduced
only into the bosonic part of the Gaussian action (\ref{eq:S0Phi}).
Specifically, we introduce the cutoff via the following substitution
in the second line of Eq.~(\ref{eq:S0Phi}),
\begin{equation}
 U^{-1} \rightarrow  U^{-1} + R_{\Lambda} (  i  \bar{\omega}  ),
 \label{eq:UR} 
\end{equation}
where
 \begin{equation}
R_{\Lambda} ( i \bar{\omega}  ) = \frac{ \Lambda}{\pi \Delta^2   }  
  R ( | \bar{\omega} | / \Lambda) ,
 \label{eq:RRdef} 
\end{equation}
and the function $R (x )$ is defined by\cite{Litim01}
 \begin{equation}
R (x) = (1-x) \Theta (1-x ).
 \label{eq:Litim} 
\end{equation} 
Our implementation of the FRG  method is therefore different from previous purely fermionic FRG
studies of the AIM, \cite{Hedden04,Karrasch08} where
no Hubbard-Stratonovich fields were introduced.
The advantages of our scheme are that the resulting FRG flow equations 
can be analyzed with moderate numerical effort and
yield a quantitatively accurate description of both the
low-energy and high-energy features (such as the broadened Hubbard peaks)
of the spectral function.
Neglecting vertex corrections, the FRG flow 
of the cutoff-dependent self-energy reads
\begin{eqnarray} 
\partial_{\Lambda} \Sigma_{\Lambda}(i\omega) & = &
\int_{\bar{\omega}} 
\dot{F}_{\Lambda} (i\bar{\omega})
G_{\Lambda} (i\omega-i  \bar{\omega}),
 \label{eq:selfflow}
 \end{eqnarray}
where the flowing fermionic propagator $G_{\Lambda} (i\omega)$
depends again on the flowing self-energy,
 \begin{equation}
 G_{\Lambda} ( i \omega ) = \frac{1}{ i \omega + i \Delta {\, \rm sign \,} \omega 
- \Sigma_{\Lambda} ( i \omega )},
 \label{eq:Dysonflow}
 \end{equation}
and the bosonic single-scale propagator is given by
\begin{equation}
 \dot{F}_{\Lambda} ( i \bar{\omega} ) = [ - \partial_{\Lambda}
 R_{\Lambda} ( i \bar{\omega}  ) ] [ F_{\Lambda} ( i \bar{\omega} ) ]^2,
 \label{eq:dotFbot} 
\end{equation}
with the bosonic propagator given by
\begin{eqnarray}
 F_{\Lambda} ( i \bar{\omega} ) & = &  
 \frac{1}{
 U^{-1} +  R_{\Lambda}   ( i  \bar{\omega}  )     -  \Pi_{\Lambda} ( i \bar{\omega} )
 }.
 \label{eq:Fbot}
 \end{eqnarray}
Here, $\Pi_{\Lambda} ( i \bar{\omega} )$ represents the flowing irreducible
spin susceptibility, which can be related 
to the fermionic Green function, in the absence of vertex corrections, 
by means of the following (skeleton) equation, \cite{Bartosch09}
\begin{eqnarray}
 \Pi_{\Lambda} ( i \bar{\omega} )  & = &  - \int_{\omega}
 G_{\Lambda} ( i \omega ) G_{\Lambda} 
( i \omega - i \bar{\omega} ) \; .
 \label{eq:skeleton2}
\end{eqnarray}
Substituting Eqs.~(\ref{eq:Dysonflow})--(\ref{eq:skeleton2})
into Eq.~(\ref{eq:selfflow}) we obtain a closed integro-differential equation
for the flowing self-energy $\Sigma_{\Lambda} ( i \omega )$, 
which must be integrated from the initial scale $\Lambda_0 \to \infty$, with 
initial condition $\Sigma_{\Lambda_0} ( i \omega ) \equiv 0$, down to $\Lambda = 0$. 
The physical self-energy is then given by the limit $ \Sigma ( i \omega ) = \lim_{\Lambda \to 0} \Sigma_{\Lambda} ( i \omega )$.

Before discussing the details of our numerical approach to Eq.~(\ref{eq:selfflow}),
it is worthwhile to remark that 
in Ref.~[\onlinecite{Bartosch09}] the corresponding flow equation for $\Sigma_{\Lambda} ( i \omega )$ 
was treated in a very crude approximation, namely by expanding all quantities
appearing in Eq.~(\ref{eq:selfflow}) to linear order in frequency.
As a consequence, only the low-energy properties of the spectrum, characterized by the $\omega = 0$ quasiparticle peak,
were accessible. Moreover, the results were obtained only in the zero-temperature limit.
Here, on the other hand, we do not rely on any low-energy expansion, so that we can
calculate the spectral function at all energy scales and describe the 
transfer of spectral weight, in the strong-coupling regime, from the Kondo peak to the high-energy Hubbard bands.

\section{Numerical solution of the FRG equation for the self-energy}

\subsection{Zero temperature}

In the zero temperature limit all the sums over Matsubara frequencies are replaced by integrals, 
$\beta^{-1} \sum_{\omega} \to \int \frac{d\omega}{2\pi}$. The flow equation for $\Sigma_\Lambda(i\omega)$ 
reduces therefore to 
an integro-differential equation, which requires an artificial discretization of the
imaginary axis in order to be solved numerically. More precisely, our numerical approach to
Eq.~(\ref{eq:selfflow}) consists in replacing the flowing self-energy $\Sigma_\Lambda(i\omega)$, 
defined as a function of the continuous parameter $\omega$, 
by a finite set of flowing couplings $\Sigma_\Lambda(i\omega_n)$ defined on a discrete mesh of frequencies $\omega_n$.

The choice of a specific parametrization of the Matsubara axis, at zero temperature, is in principle 
arbitrary; however, one has to carefully verify that the numerical results are actually independent of such a parametrization. 
In other words, the results should be numerically stable with respect to the choice of the discretization procedure.
To achieve this goal we consider the following geometric mesh\cite{Karrasch08} of frequencies,
\begin{equation}
 \omega_n = \omega_{\rm min} \frac{ a^n -1 }{ a-1 }, \qquad n=1,\ldots,N,
\end{equation}
where the free parameters $a>1$ and $\omega_{\rm min}>0$ define the spacing of the frequencies, 
while $N$ is the total number of frequencies to be used (note that in the particle-hole symmetric case we have
$\Sigma_\Lambda(i\omega)= -\Sigma_\Lambda(-i\omega)$, so that we do not need independent couplings for the negative frequencies).  
As thoroughly discussed in Ref.~[\onlinecite{Karrasch08}], the unequal spacing of such a mesh allows indeed to resolve with
great accuracy the low-energy regime of the spectrum, which is known to exhibit the sharp Kondo resonance at $\omega =0$, 
while covering, at the same time, a sufficiently wide range of frequencies in order to rule out the spurious effects
of a finite frequency cutoff $\omega_{\rm max} = \omega_{N}$. 
More specifically, one has to simultaneously satisfy, for all the values of $U$ under consideration, the following conditions,
\begin{eqnarray}
 \omega_{\rm min} & \ll & Z(U) \Delta, \\
 \omega_{\rm max} & \gg & \max (\Delta, U),
\end{eqnarray}
where 
\begin{equation}
Z(U) = \left(1 - \left. \frac{\partial \Sigma (i\omega)}{ \partial (i\omega) } \right|_{ \omega =0} \right)^{-1}
\end{equation}
is the quasiparticle residue  
(proportional to the width of the Kondo peak),
and afterwards verify that the results are numerically stable upon a further increase in both
the spanned frequency range $[\omega_{\rm min}, \omega_{\rm max}]$ and the number $N$ of sampling frequencies.
A typical choice of the discretization parameters is, e.g.,
\begin{equation}
\omega_{\rm min} = 10^{-6} \Delta, \quad a=1.06, \quad N=400, 
\end{equation}
which allows to solve the flow equations with a relatively small computational effort 
up to the strong-coupling regime $U \lesssim 8 \pi \Delta$.

The resulting system of $N$ coupled differential equations for the flowing couplings $\Sigma_\Lambda(i\omega_n)$
is solved by means of refined Runge-Kutta routines, using a linear interpolation procedure in order to evaluate the 
Green function in Eqs.~(\ref{eq:selfflow}) and (\ref{eq:skeleton2}) at frequencies that do not belong to the geometric mesh 
(i.e., when $\omega_{\ell} - \bar{\omega}_{\ell'} \notin \{ \omega_n \}$).  

\subsection{Finite temperatures}

At finite temperatures the Matsubara space is intrinsically discrete, with fermionic and bosonic
frequencies given by
\begin{equation}
 \omega_n = (2 n +1)\pi/\beta, \qquad
 \bar{\omega}_n = 2 n \pi/\beta, 
 \label{eq:Matsubara_f}
\end{equation}
so that the flowing self-energy itself is already defined as a countable set of couplings $\Sigma_\Lambda(i\omega_n)$.
Eq.~(\ref{eq:selfflow}) is therefore no longer a proper integro-differential equation, as in the zero temperature limit,
but consists rather of an infinite set of coupled ordinary differential equations, 
which can be solved numerically by keeping only the first $N$ Matsubara frequencies.

In contrast to the zero temperature case, it is crucial to observe that now the spacing between
consecutive points of the mesh is constant, its value $\delta \omega = 2\pi T$ being dictated by the 
actual value of the temperature $T$. This fact simplifies the implementation of the numerical routines, since there is no need
for any interpolating procedure; on the other hand, however, it limits the possibility of studying 
arbitrarily small temperatures. In fact, in order to obtain numerically stable results using the natural
Matsubara discretization defined in Eq.~(\ref{eq:Matsubara_f}), 
one typically needs $N \gtrsim 10^3$ for temperatures $T/\Delta \lesssim 0.1$, which already corresponds to
a remarkable computational effort. In other words, although smaller values of $T/\Delta$ are still accessible from the numerical point 
of view, the computational time required becomes comparable to that of fully numerical methods (e.g. the NRG), 
making the present method numerically unworthy.

\subsection{Analytical continuation}

The solution of our flow equation gives, by construction, the physical
self-energy along the imaginary axis, $ \Sigma ( i \omega ) = \Sigma_{\Lambda=0} ( i \omega )$. 
In order to access real frequency properties, such as the spectral function 
\begin{equation}
 A(\omega) = -\frac{1}{\pi} {\rm Im \,} G(i\omega \to \omega + i0),
\end{equation}
we need to analytically continue our results for either $G(i\omega)$ or $\Sigma(i\omega)$ onto the real frequency axis.
Since our numerical solutions are not affected by statistical errors, 
we perform the analytical continuation using the Pad\'e approximation, \cite{Pade77} which we found to
give more stable results if directly applied to the self-energy. In other words, we first evaluate the real
frequency self-energy, $ \Sigma(\omega) = \Sigma( i\omega \to \omega + i0)$, and afterwards calculate the
corresponding Green function using the Dyson equation. The frequency dependence of the self-energy
is indeed much more sensitive to the interaction-induced features of the model than the Green function itself, whose non-trivial
behavior is masked by the asymptotic form of the non-interacting propagator.

\section{Numerical Renormalization Group at Finite Temperatures}

The numerical renormalization group method \cite{Wilson75} was
first applied to the Anderson impurity model \cite{Krishna-murthy80} shortly
after its introduction in 1975. Originally designed to
calculate thermodynamic quantities at very low temperatures,
the NRG has been extended over the last two
decades to calculate more complex observables, such as
the single particle spectral function investigated here
(for a recent review covering many extensions and applications of the  
NRG, see Ref.~[\onlinecite{Bulla08}]).
We briefly mention several crucial developments which lead to  
improving the accuracy
of dynamical correlation functions calculated
by NRG: the $z$-trick averaging \cite{Yoshida90} which
improves the resolution at frequencies close to the conduction-band edge; 
the introduction of the reduced density
matrix \cite{Hofstetter00} in order to account for the correct low-temperature state  
when
calculating the spectrum at higher frequencies;
and finally, a recently developed smart choice of the basis for the  
full NRG Fock space, \cite{Anders05}
which was designed to calculate the real-time evolution
after a quantum quench, and successively has
been applied to improve the accuracy of spectral
functions for the single impurity Anderson model \cite{Peters06} by avoiding
overcounting of spectral transitions within the NRG.

The NRG data presented here were obtained by the full density matrix  
approach, \cite{Weichselbaum07}
which gives access to reliable finite-temperature spectra for  
frequencies as low as $\omega \approx T/5$.
In combination with the self-energy trick \cite{Bulla98} and an average
over $N_{z} = 32$ slightly different discretizations of the
conduction band \cite{Yoshida90} we obtain high-quality NRG spectra,
which serve as a benchmark for the newly developed FRG
scheme discussed above. A detailed description of the
NRG method is beyond the scope of this publication; for details
we refer the interested reader to Ref.~[\onlinecite{Weichselbaum07}] and references therein.

 \begin{figure}[!tb]
 \centering
 \includegraphics*[width=82mm]{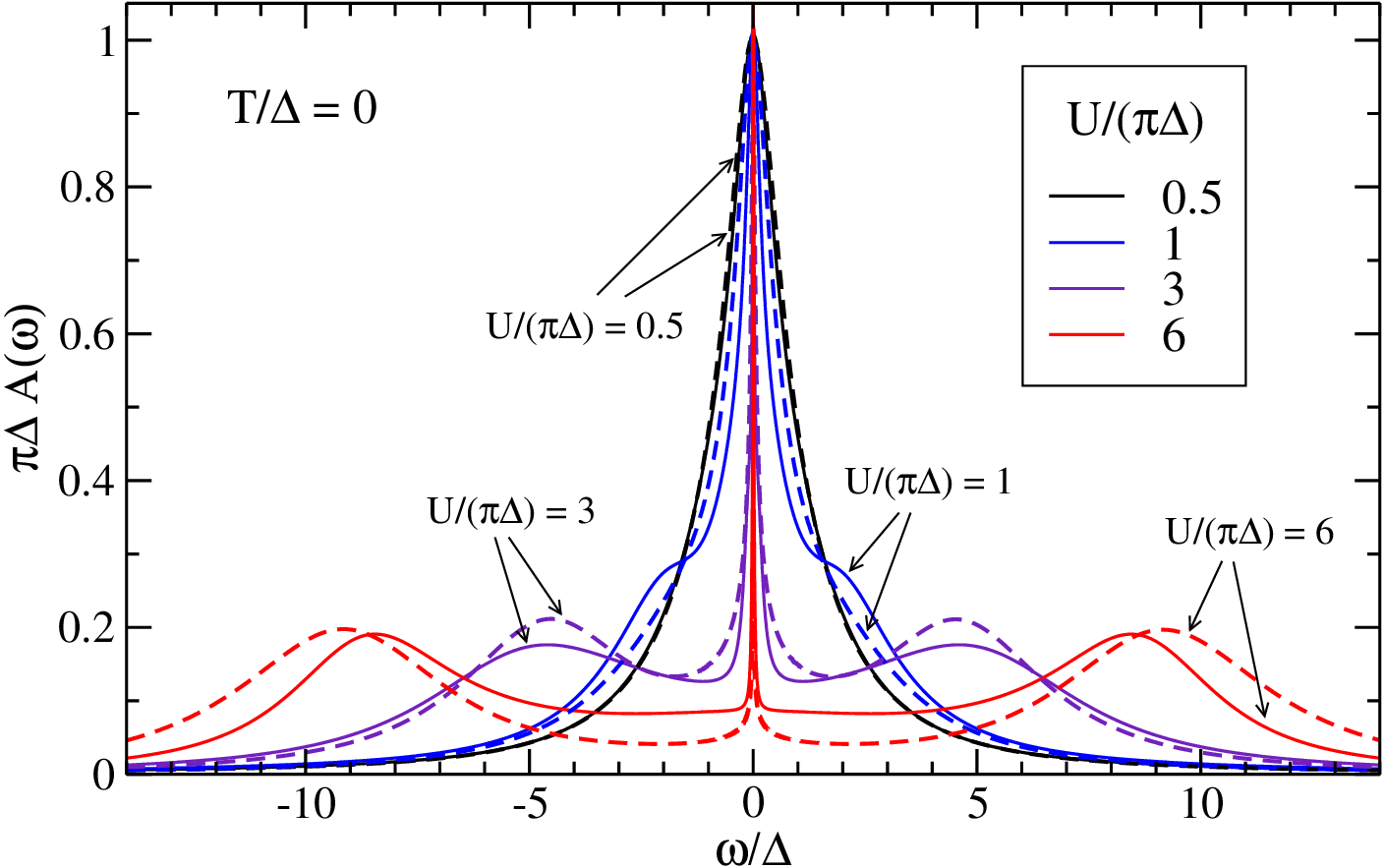}
% \vskip2mm
 \includegraphics*[width=82mm]{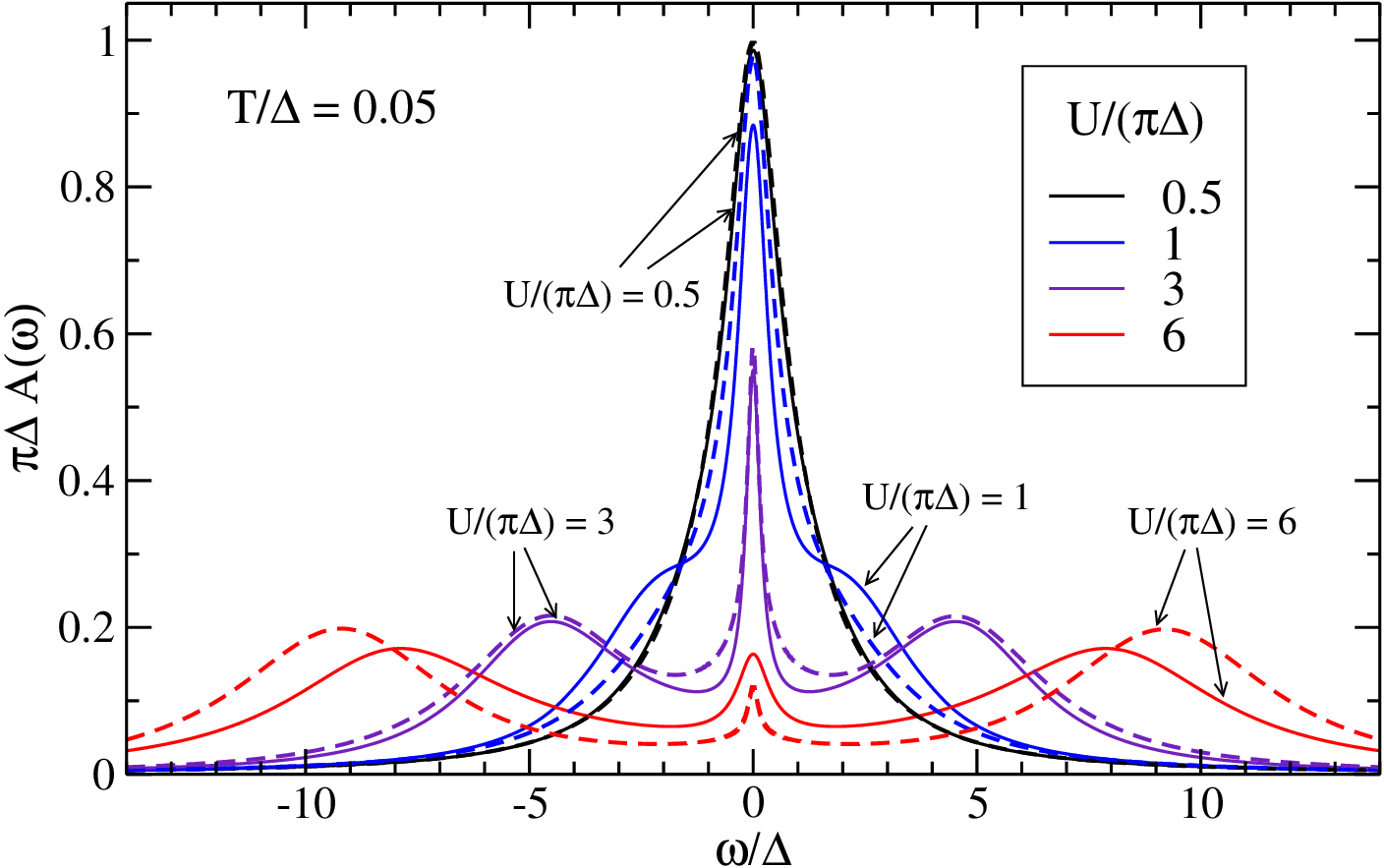}
% \vskip2mm
 \includegraphics*[width=82mm]{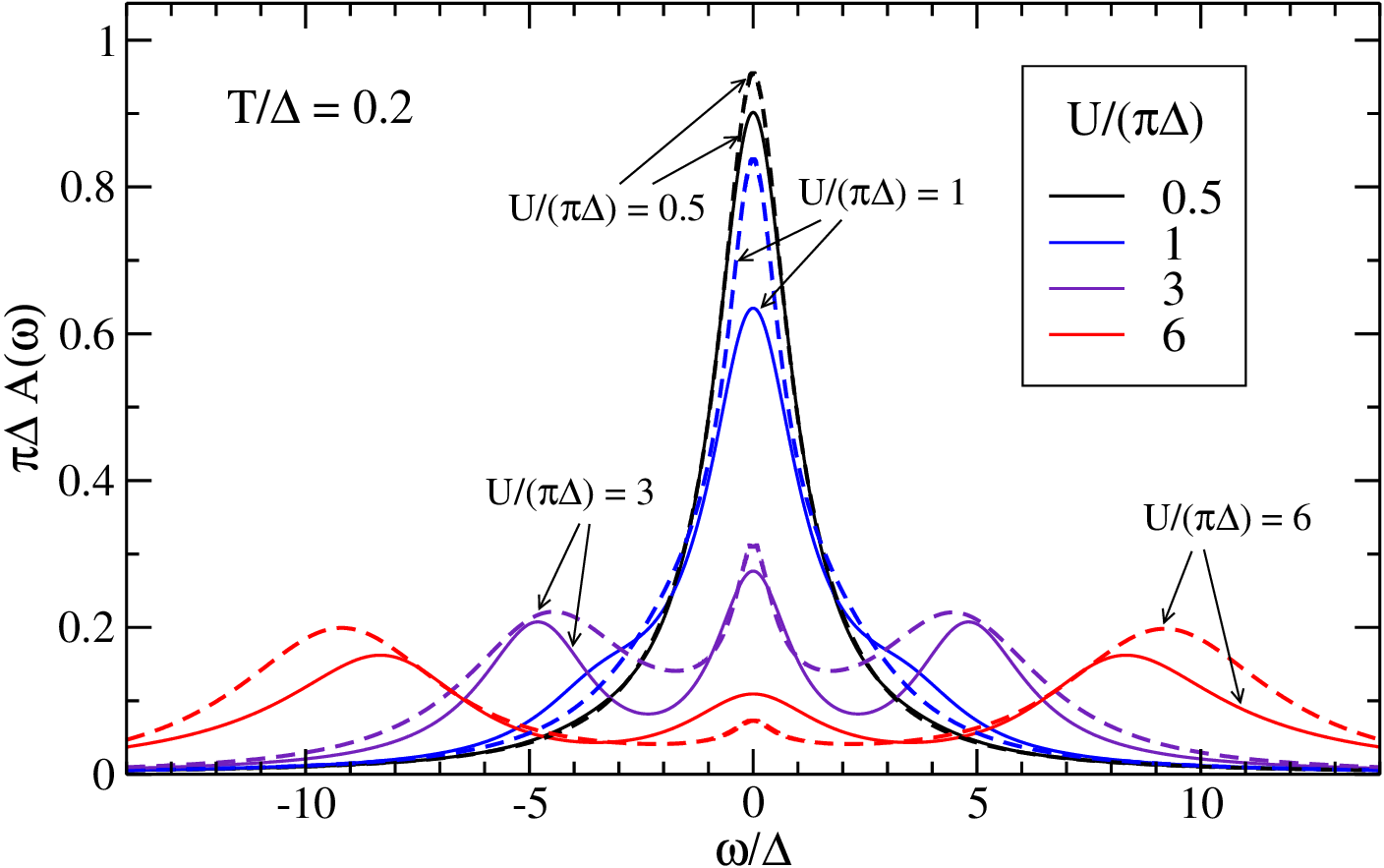}
 \caption{%
 (Color online)
 Spectral function of the symmetric AIM evaluated with the FRG (solid lines) and the NRG (dashed lines) for different interactions and temperatures.
 From top to bottom the temperature increases from $T/\Delta = 0$ to $T/\Delta = 0.2$, as indicated.
 }
 \label{fig:spectra_1}
 \end{figure}

\section{Results}

 \begin{figure}[!tb]
 \centering
 \includegraphics*[width=82mm]{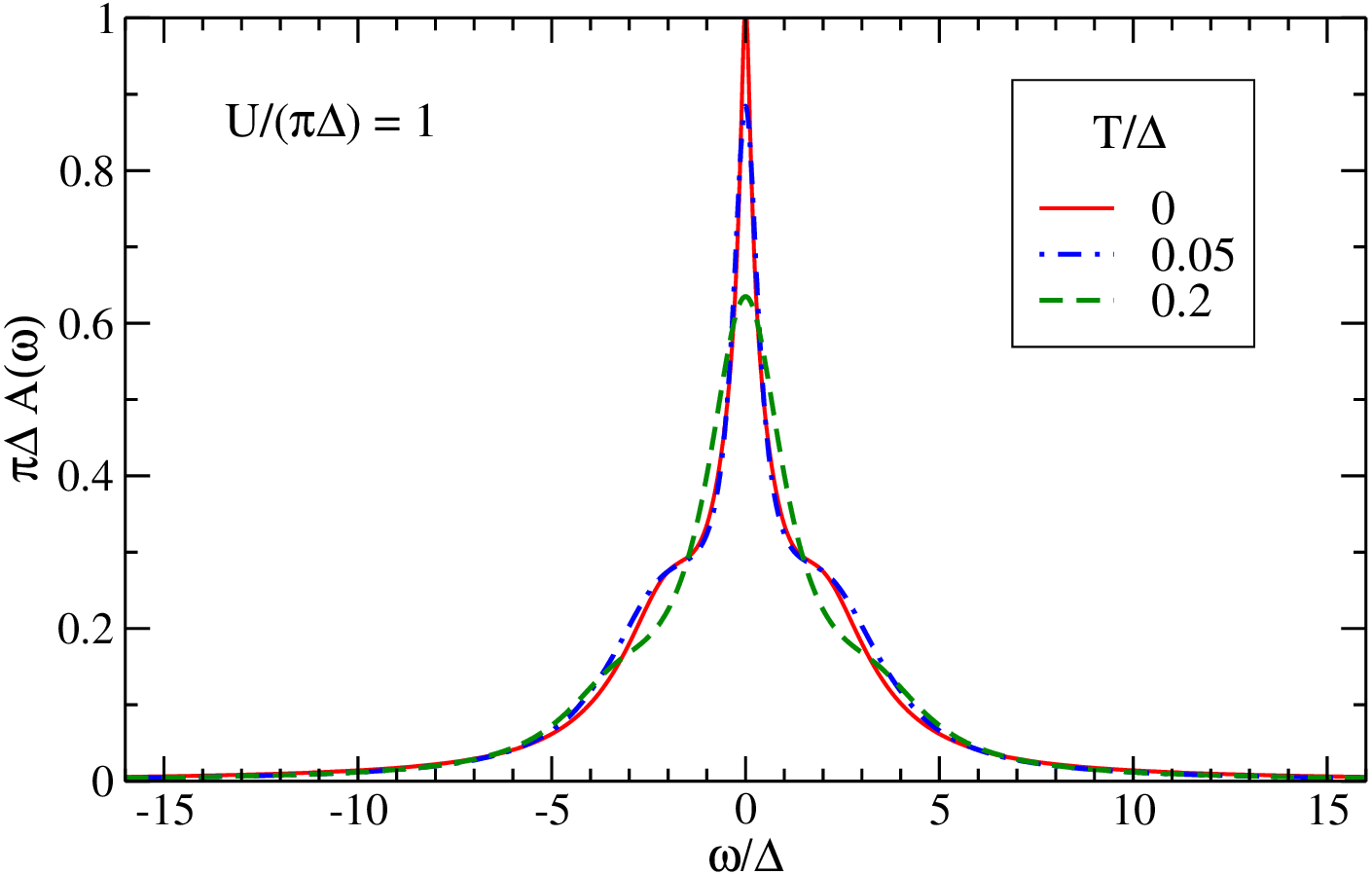}
%\hskip1.5ex
 \includegraphics*[width=82mm]{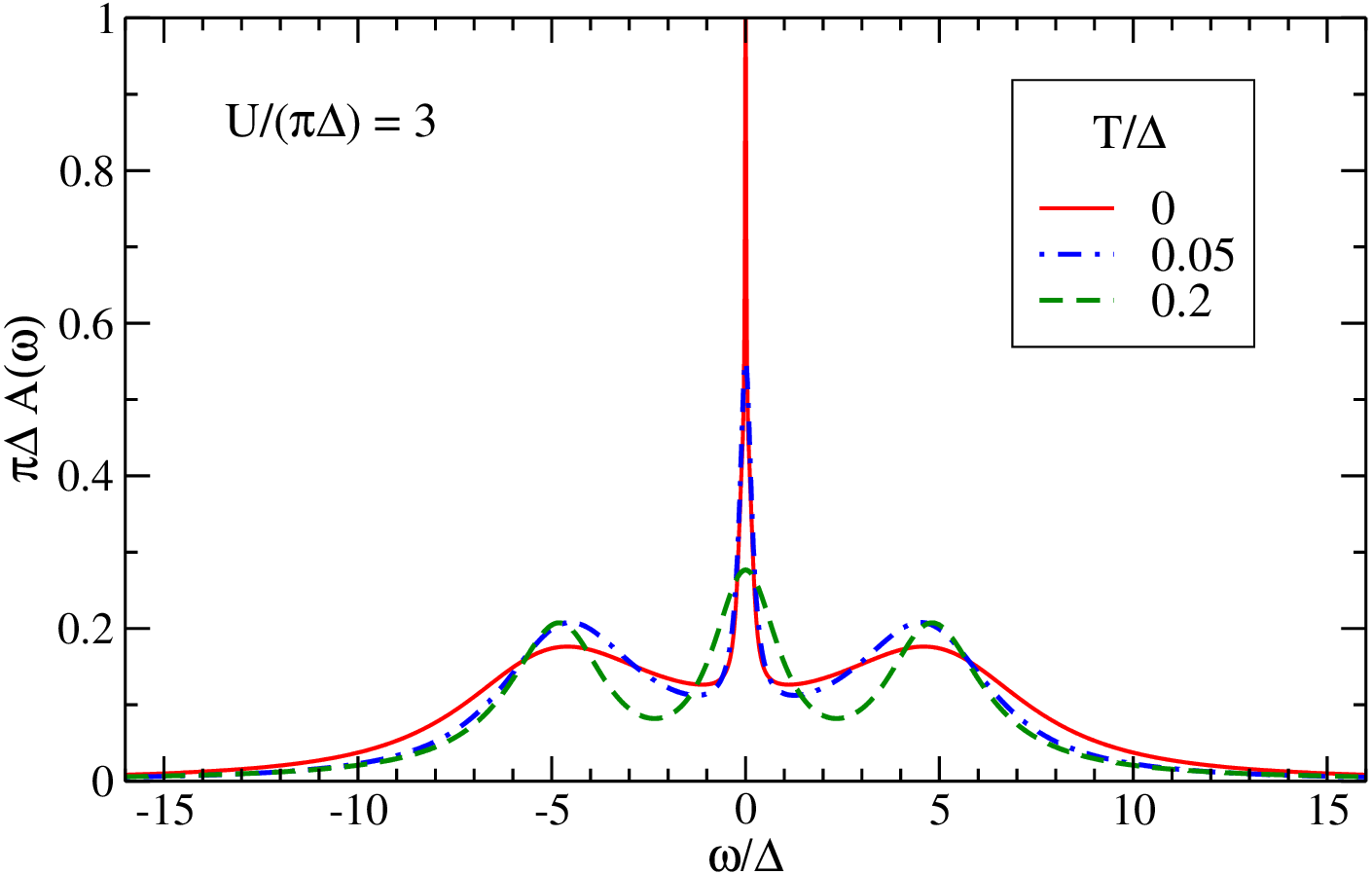}
%\hskip1.5ex
 \includegraphics*[width=82mm]{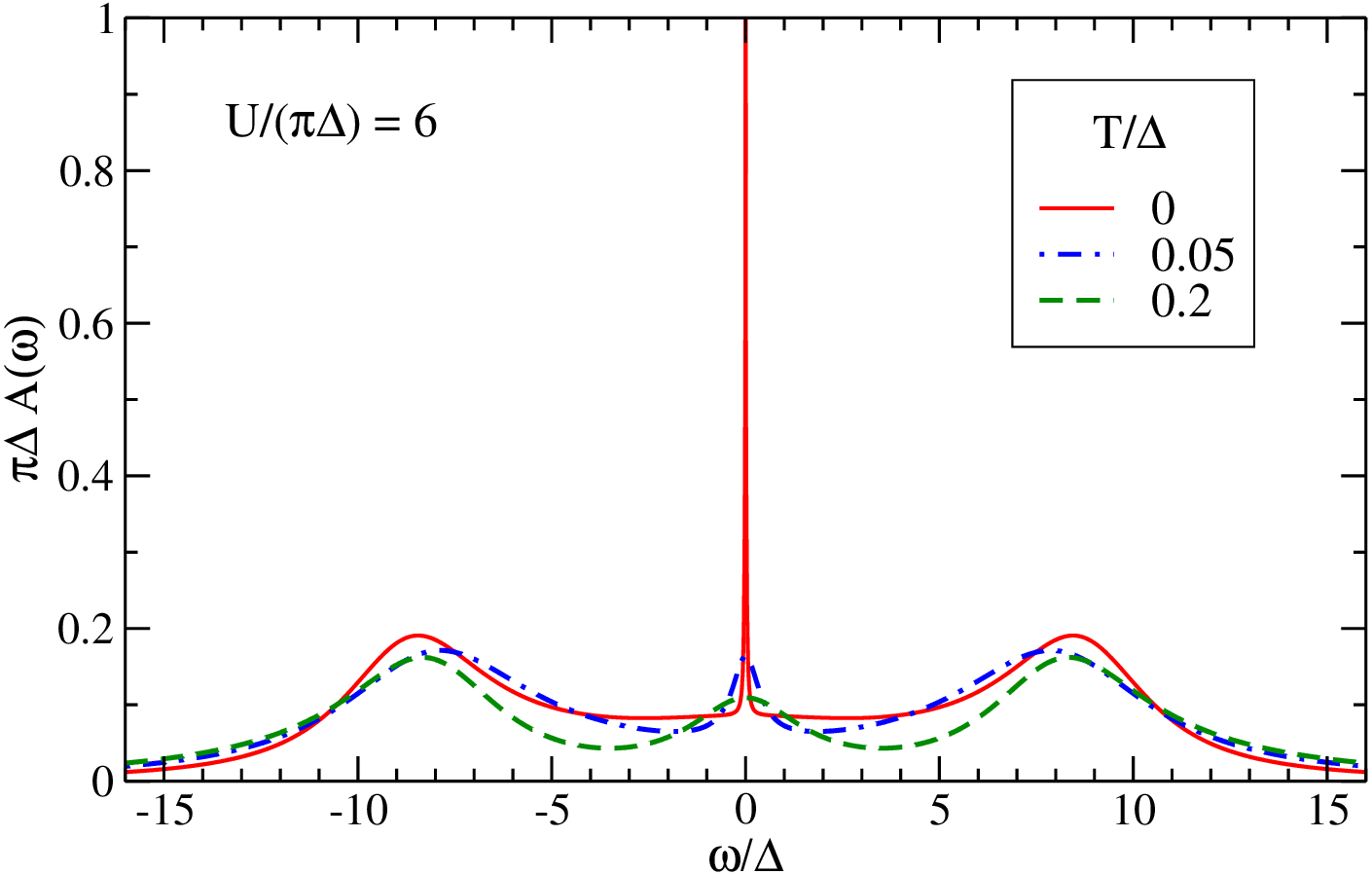}
 \caption{%
 (Color online)
 FRG spectral function for different interactions and temperatures.
 From top to bottom the interaction increases from $U/(\pi\Delta) = 1$ to $U/(\pi\Delta) = 6$, as indicated.
 }
 \label{fig:spectra_2}
 \end{figure}

Our results for the spectral function at temperatures $T/\Delta = 0$, 0.05 and 0.2 are plotted in Fig.~\ref{fig:spectra_1},
where in each panel we have considered several values of the interaction strength, ranging from the weak to the strong coupling regime.
The zero temperature line-shapes are in good agreement with the NRG data in the weak and strong coupling regimes, while
in the intermediate coupling range $U/(\pi\Delta) \approx 1$ our method somewhat overestimates the role of the interaction, enhancing the transfer 
of spectral weight from the Kondo resonance to the Hubbard bands. It is worthwhile to observe that for large values of the interaction the
peaks of the Hubbard bands are located approximately at $\pm U/2$ and their width is of order $2\Delta$, in good agreement with the known
strong-coupling limit results. \cite{Hewson93} 
As the temperature increases, the Kondo resonance peak becomes more and more broadened, as expected when the
temperature approaches the Kondo scale $T_K$. However, we notice that at small and intermediate couplings 
this effect is more enhanced in the FRG approach, in comparison to the NRG data.  
On the other hand, the position and the shape of the Hubbard bands remain
essentially unchanged, being related to an energy scale much larger than the temperatures under consideration.
This behavior is more clearly shown in Fig.~\ref{fig:spectra_2}, where the FRG results are now plotted, in each panel, 
for a fixed value of the interaction and increasing temperatures.

 \begin{figure}[!tb]
 \centering
 \includegraphics*[width=82mm]{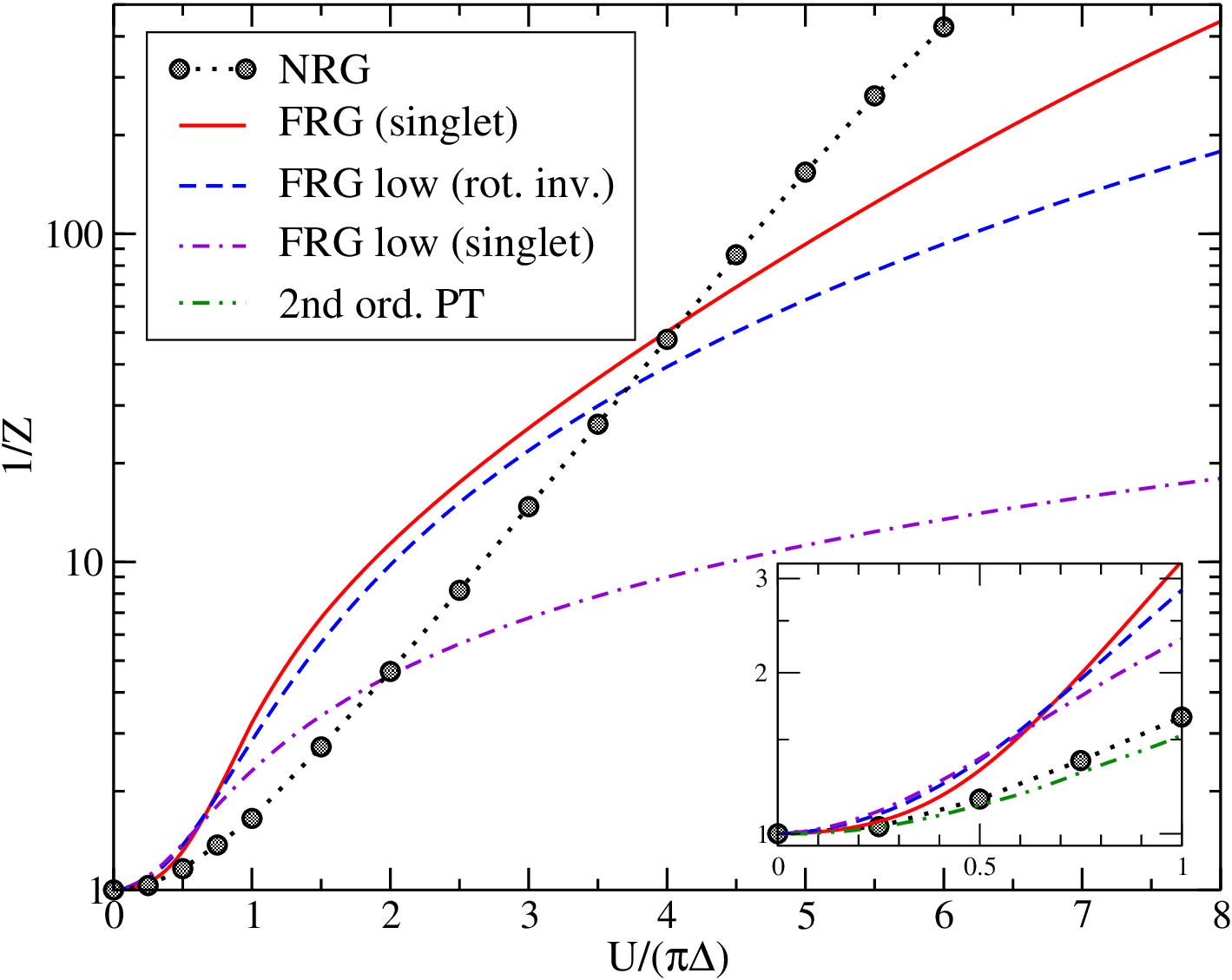}
 \caption{%
 (Color online)
 Inverse quasiparticle weight $Z^{-1}$ at zero temperature on a 
 logarithmic scale, as a function of $U / ( \pi \Delta)$.
 The inset shows the weak-coupling behavior in the regime $0 < U / ( \pi \Delta) < 1$
 on a larger scale. The present FRG method corresponds to the solid curve, the dashed and dot-dashed curves are
 the FRG results of Ref.~[\onlinecite{Bartosch09}], the black circles represent the NRG data 
 and the dot-dot-dashed curve in the inset is the second order perturbation theory.
 }
 \label{fig:Zplot}
 \end{figure}

Finally, in Fig.~\ref{fig:Zplot} we show the inverse quasiparticle weight at $T=0$ 
as a function of the interaction. Unfortunately, in contrast to the NRG results, at strong coupling
our FRG method is not able to reproduce the exponential 
narrowing of the quasiparticle peak predicted by the Bethe Ansatz. \cite{Tsvelick83}
However, the numerical estimate of $1/Z$ for large values of $U/(\pi\Delta)$
is improved compared to the previous FRG results presented in Ref.~[\onlinecite{Bartosch09}]. 
Fig.~\ref{fig:Zplot} shows also clearly the above mentioned overestimation of the interaction in 
the intermediate coupling regime, characterized by a stronger renormalization of the quasiparticle peak.

\section{Conclusions}

In the present work we have calculated the spectral function of the symmetric Anderson impurity model
at zero and finite temperatures using functional and numerical renormalization group methods. 
In particular, we took advantage of our high quality NRG data in order to test the validity of
the FRG truncation scheme proposed in Ref.~[\onlinecite{Bartosch09}], where the on-site interaction is
decoupled in the spin-singlet particle-hole channel by means of a bosonic Hubbard-Stratonovich field 
and an infrared cutoff is explicitly introduced only into the bosonic propagator.
Following Ref.~[\onlinecite{Bartosch09}], we neglected the FRG flow of vertex functions in order to
truncate the infinite hierarchy of FRG flow equations and obtain a closed integro-differential equation for the
flowing self-energy $\Sigma_\Lambda(i\omega)$. In contrast to Ref.~[\onlinecite{Bartosch09}], however, 
where only the low-energy expansion of the self-energy was 
taken into account, we solved numerically the integro-differential equation for $\Sigma_\Lambda(i\omega)$ 
keeping the complete frequency structure. This has allowed us to calculate the spectral function at all energy scales,
including the high-energy Hubbard bands.

Comparing the FRG results to the NRG data, we found that our truncation scheme
gives quantitatively good results for the spectral line-shape in the weak and strong coupling regimes, 
both at zero and finite temperatures, although
it somewhat overestimates the 
effects of the interaction at intermediate couplings $U/(\pi\Delta) \approx 1$.
Most important, in contrast  
to purely fermionic FRG approaches \cite{Hedden04,Karrasch08} where no Hubbard-Stratonovich fields are 
introduced, our method gives an accurate description of both the low-energy 
and high-energy features of the spectral function, including the correct position and width 
of the high-energy Hubbard bands in the strong coupling regime.

Unfortunately, our FRG truncation scheme is still not able to correctly reproduce, 
at zero temperature, the exponential narrowing of the Kondo peak
in the strong coupling regime. Most likely, such a non-perturbative effect requires exact symmetry 
relations, e.g.\ Ward identities, \cite{Kopietz10b} to be enforced throughout the integration of the FRG flow.

\section*{ACKNOWLEDGMENTS}
We thank Volker Meden for useful comments.
This work was supported by the DFG 
via Sonderforschungsbereich 
SFB/TRR 49 and Forschergruppe FOR 723.
We also acknowledge financial support by
the DAAD/CAPES PROBRAL-program. 
The work by A.I. and P.K. was partially carried out at the
International Center for Condensed Matter Physics (ICCMP) at the University
of Bras\'ilia, Brazil. We thank the director of the ICCMP, Alvaro Ferraz, for his 
hospitality.

% Hofstetter's group  SFB transregio acknowledgments  

\end{document}